\documentclass[12pt,longbibliography]{iopart}

\usepackage{iopams}
\usepackage{graphicx}
\usepackage{cite}
\usepackage{footmisc}
\usepackage{url}
\begin{document}
\maketitle\pagestyle{plain}

\title{Uncovering double-stripe and plaquette antiferromagnetic states in the one-band Hubbard model on a frustrated square lattice}

\author{Ze Ruan\footnote{\label{contribution}These authors contribute equally to the paper}, Xiu-Cai Jiang\footref{contribution}, Ze-Yi Song, Yu-Zhong Zhang}

\address{Shanghai Key Laboratory of Special Artificial Microstructure Materials and Technology, School of Physics Science and engineering, Tongji University, Shanghai 200092, P.R. China}
\ead{\mailto{zysong@tongji.edu.cn},\mailto{yzzhang@tongji.edu.cn}}

\date{\today}

\begin{abstract}
Groundstate magnetism of the one-band Hubbard model on the frustrated square lattice where both nearest-neighbour $t_1$ and next-nearest-neighbour $t_2$ hoppings are considered at half-filling are revisited within mean field approximation. Two new magnetic phases are detected at intermediate strength of Hubbard $U$ and relative strong frustration of $t_2/t_1$, named double-stripe and plaquette antiferromagnetic states, both of which are metallic and stable even at finite temperature and electron doping. The nature of the phase transitions between different phases and the properties of the two new states are analyzed in detail. Our results of various magnetic states emerging from geometric frustration in the minimal model suggests that distinct antiferromagnetism observed experimentally in the parent states of two high-T$_c$ superconducting families, i.e., cuprates and iron-based superconductors, may be understood from a unified microscopic origin, irrespective of orbital degrees of freedom, or hoppings further than next-nearest neighbour, etc.
\end{abstract}

%
%
%
%
%

\section{Introduction}

Antiferromagnetism has attracted tremendous interests due to the fact that the parent states of two high-T$_c$ superconducting families, such as cuprates and iron-based superconductors, are either antiferromagnets or paramagnets with antiferromagnetic fluctuations~\cite{Keimer2015,Dai2015RMP}. And it is commonly accepted that fully understanding of antiferromagnetism may provide a key to unravel the mechanism of high-T$_c$ superconductivity~\cite{Lee2006RMP,Scalapino2012RMP}. While the two families can be generally modeled on frustrated square lattices formed by copper and iron ions, respectively, two fundamental differences between these two systems hinder the development of a unified theory for high-T$_c$ superconductivity and corresponding antiferromagnetism.

The first is that, contrast to only one band crossing the Fermi level in conventional bulk cuprates with single copper oxide layer, five $3d$ orbitals are all active in the low-energy region of iron-based superconductors. Recently, this difference seems to be reconciled after comprehensive understanding of the orbital selectivity~\cite{GeorgesARCMP2013}. It is found that the Hund's coupling suppresses the inter-orbital charge fluctuations~\cite{Medici2011PRB,Song2017PRB}, leading to decoupling of the orbitals and separated phase transitions in each orbital as a function of electronic interacting strength or doping, etc~\cite{Medici2014PRL,Yu2013PRL}. Therefore, in order to gain a deep insight into the origin of all the phases detected in iron-based superconductors, study of a complicated multi-orbital system can be first decomposed into investigations of multiple single-orbital models. And then, details of the whole phase diagram which is analogous to that of cuprates can be qualitatively derived from the interplay among distinct phases of different orbitals. Explicitly speaking, results on a proper one-band model can be used as a starting point to understand physical properties of both high-T$_c$ superconducting families.

The other difference is that, while the parent states of high-T$_c$ cuprates are of checkerboard antiferromagnetic (CAF) insulator, various antiferromagnetic patterns were observed experimentally in the mother compounds of iron-based superconductors, such as nearly degenerate double-stripe (DAF) and plaquette antiferromagnetic (PAF) order in FeTe~\cite{Li2009PRB,Ma2009PRL,Tam2019PRB,Ducatman2012PRL}, pair-checkerboard antiferromagnetic (PCAF) order in monolayer FeSe thin film grown on SrTiO$_3$(001)~\cite{Zhou2018PRL,Cao2015PRB}, molecular-intercalated FeSe~\cite{Taylor2013PRB}, and A$_x$Fe$_{2-y}$Se$_2$~\cite{Taylor2012PRB}, as well as stripe-type antiferromagnetic (SAF) order in most others~\cite{nature2008Cruz,Huang2008PRL,PRB2009Li}. Although evolution from CAF to SAF orders can be explained within a Heisenberg model on the frustrated square lattice, which can be viewed as the strong coupling limit of a one-band Hubbard model with both nearest- (t$_1$) and next-nearest-neighbour (t$_2$) hoppings at half-filling, it was frequently mentioned in the literature that others like DAF, PAF, PCAF can not be understood with the same model unless third-neighbour exchanges and nearest-neighbour biquadratic exchanges are additionally involved~\cite{Hu2012PRB,Glasbrenner2015NatPhy}.

However, we noticed that the PCAF order with double periodicity in one direction compared to CAF order had been theoretically proposed to appear in the phase diagram of the above Hubbard model, which is located between CAF and SAF phases since geometric frustration is the strongest in that region~\cite{Mizusaki2006PRB}. In fact, geometric frustration of interacting electronic systems can lead to a rich variety of phenomena, such as Mott metal-insulator transition~\cite{Imada1998RMP}, unconventional superconductivity~\cite{Kyung2006PRL,Nevidomskyy2008PRB,Hassan2008PRB}, and exotic quantum magnetic orderings~\cite{Misumi2016JPSJ,Zhou2017RMP,Misumi2017PRB}. Thus, interesting questions arise; whether the DAF and PAF orders can also emerge from perturbations to the macroscopic degeneracies induced by strong geometric frustration, and whether diverse antiferromagnetic states, either insulating or metallic, can be explained in a unified minimal model, i.e., the one-band Hubbard model on the frustrated square lattice with hoppings only up to next-nearest neighbour at half-filling, which covers both itinerant picture for magnetism at weak coupling limit and localized scenario at strong coupling limit?

Such a model has been extensively studied and complex phase diagrams were obtained by different non-perturbative methods, including path integral Monte Carlo (PIMC)~\cite{Mizusaki2006PRB}, variational Monte Carlo (VMC)~\cite{Tocchio2008PRB} and variational cluster approximation (VCA)~\cite{Nevidomskyy2008PRB,Yoshikawa2009PRB,Yamada2013PRB,Yamaki2013PRB,Misumi2016JPSJ}. Though controversies still remain regarding the intermediate $t_2$ region where effect of frustration becomes strong, the magnetic phases at small and large $t_2$ are qualitatively consistent among different studies. When the Coulomb interaction exceeds a threshold, the CAF state was found to be stabilized at small $t_2$, which has been directly observed in recent ultracold atom experiments~\cite{Mazurenko2017Nature,Drewes2017PRL}. Since the presence of $t_2$ yields antiferromagnetic spin exchange between next-nearest-neighbour electrons and is destructive to the CAF state where the next-nearest-neighbour spins align ferromagnetically, the increasing of t$_2$ consequently leads to a phase transition from the CAF state to a paramagnetic metal at weak onsite electron repulsion or a SAF state at large Hubbard $U$. Till now, whether competing magnetic states like DAF and PAF states would emerge in the vicinity of strongest frustration region of $t_2={t_1}/\sqrt{2}$ and intermediate value of $U$ has not yet been explored.

\begin{figure}[htbp]
 \centering
	\includegraphics[width=0.9\textwidth]{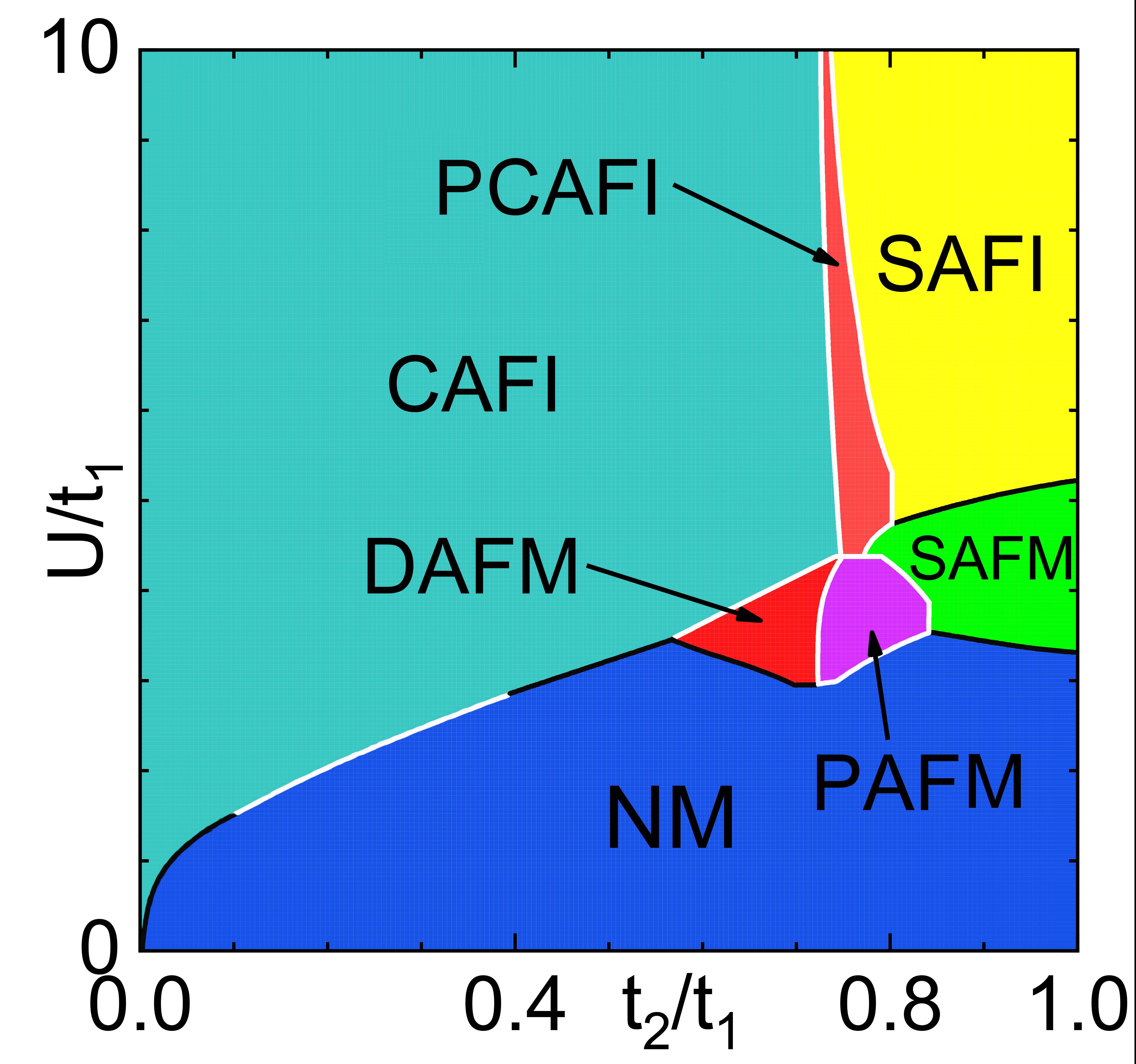}
	\caption{Phase diagram of the two dimensional Hubbard model on the square lattice with nearest-neighbour $t_1$ and next-nearest-neighbour $t_2$ hoppings at half filling and zero temperature, $U$ is the onsite Coulomb interaction. CAFI denotes the checkerboard antiferromagnetic insulator, NM the nonmagnetic metal, DAFM the double-stripe antiferromagnetic metal, PAFM the plaquette antiferromagnetic metal, PCAFI the pair-checkerboard antiferromagnetic insulator, SAFM and SAFI stand for the stripe-type antiferromagnetic metallic and insulating states, respectively. The white and black lines represent the corresponding first-order and  second-order phase transitions.}
	\label{phase_diagram}
\end{figure}

In this paper, we have reinvestigated the role of geometrical frustration in the Hubbard model on the square lattice with $t_1$ and $t_2$ at half filling and zero temperature by using mean field approximation. We only focus on the static magnetic properties as mean field approximation can not capture the quantum fluctuations. It is known that in the presence of long-range order, quantum fluctuations will be largely suppressed, and the mean field results are qualitatively reliable~\cite{Zhang2012PRB}. The calculated $U/t_1-t_2/t_1$ phase diagram is shown in Fig.~\ref{phase_diagram}. In comparison to previous theoretical results, two new antiferromagnetic states, i.e. the DAF and PAF states, are found to emerge in the region $0.56\leq{}t_2/t_1\leq0.85$ and $2.93\leq{}U/t_1\leq4.4$ as a result of the interplay and competition between kinetic energy, Coulomb repulsion and geometric frustration. The energetic stability of the DAF and PAF states over other antiferromagnetic states is further confirmed at finite temperature and finite doping. In the new phases, the system are metallic. In PAF state, additional Lifshits transition occurs. The relevance of the two new phases to real materials will be discussed.

\begin{figure}[htbp]
	\centering
	\includegraphics[width=0.9\textwidth,height=0.7\textheight]{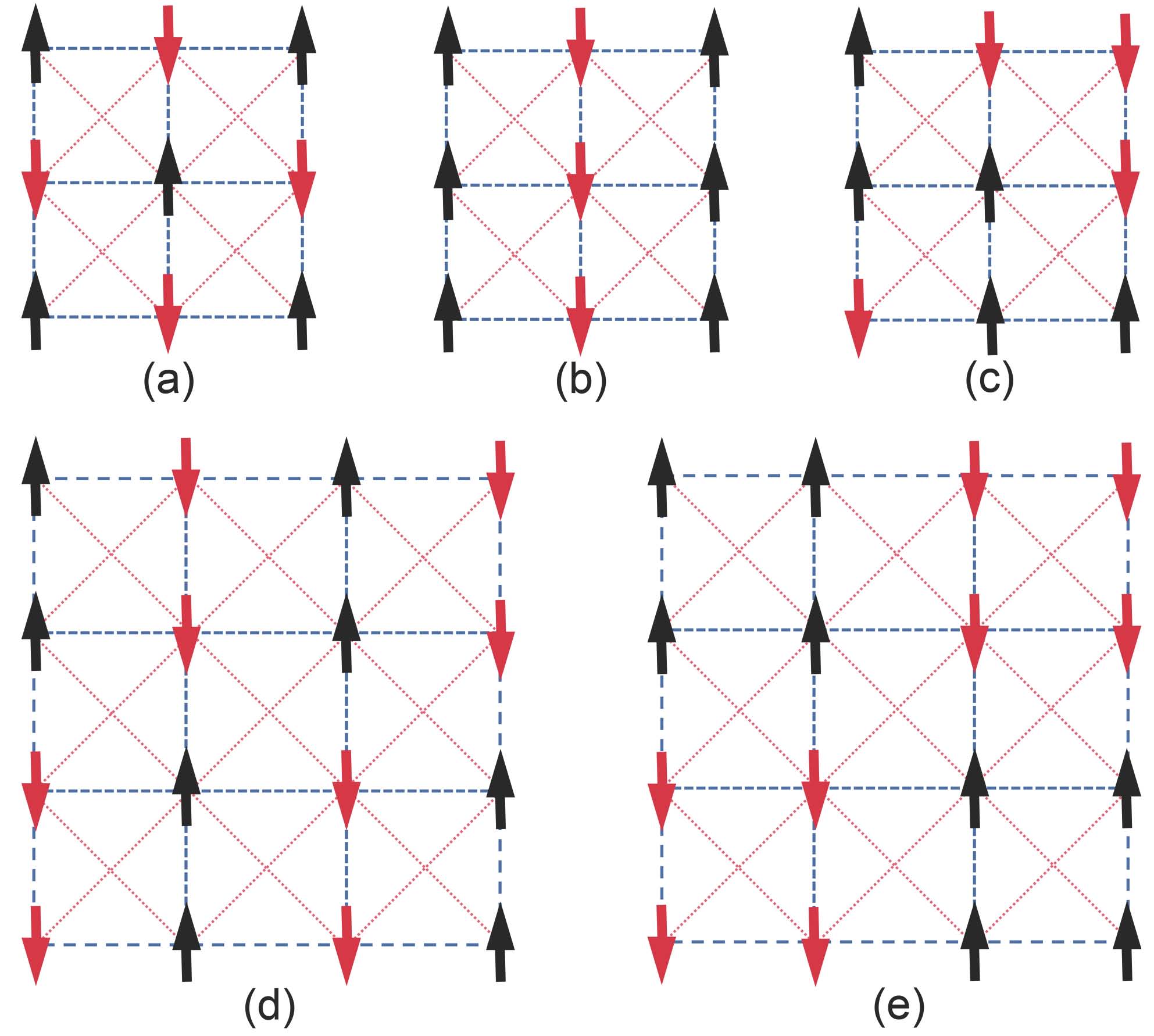}
	\caption{Spin configurations of the CAF state (a), SAF state (b), DAF state (c), PCAF state (d) and PAF state (e).}
	\label{SpinConfig}
\end{figure}

\begin{figure*}[htbp]
	\centering
	\includegraphics[width=1\textwidth]{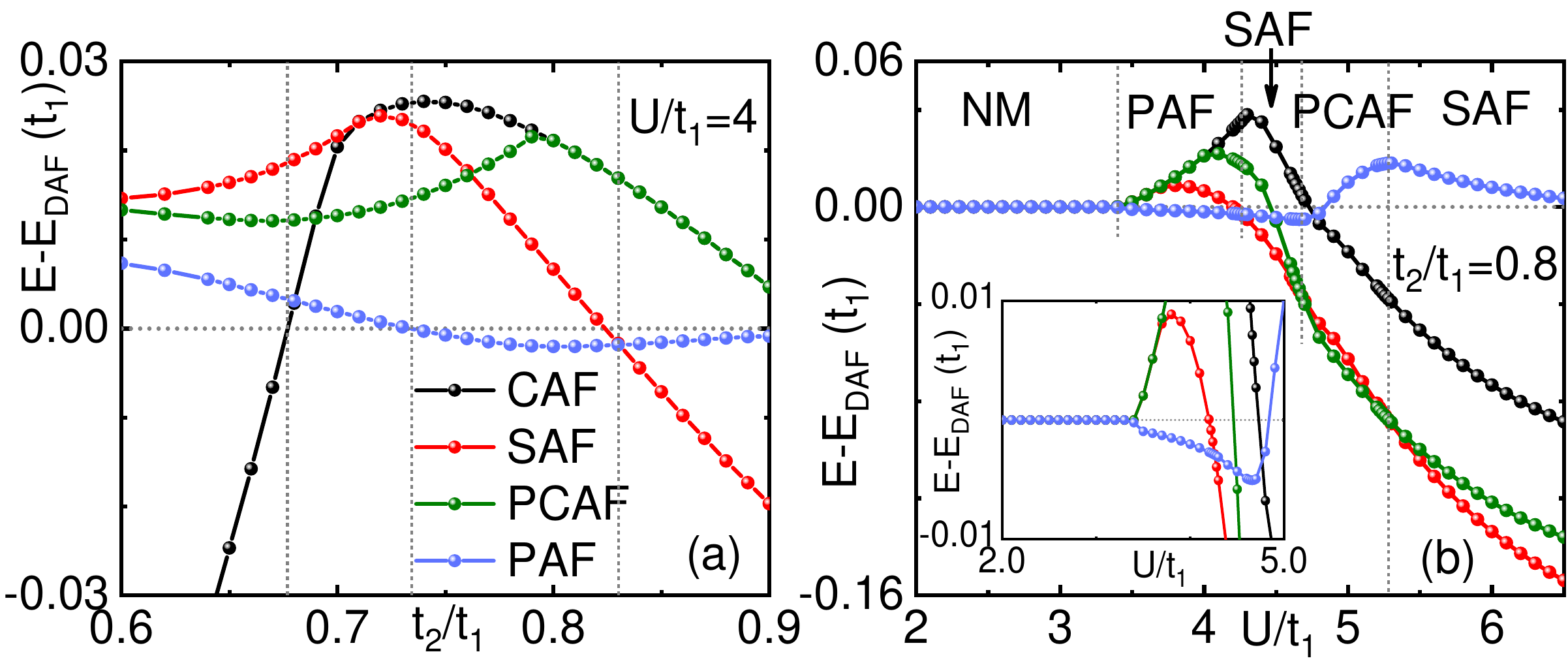}
	\caption{ (a) The total energies of the CAF, SAF, PCAF and PAF states as a function of the next-nearest-neighbour hopping $t_2$ at $U/t_1=4$, where the energy of the DAF state is set to zero. (b) At $t_2/t_1=0.8$, the change of the ground-state magnetic ordering with the increase of $U$.  When $U$ is small, the ground state of the system is a NM metal. The inset is a blowup for the switch of the ground state when $2.0<U/t_1<5.0$. }
	\label{Totalenergies}
\end{figure*}

\section{model and method}

The Hubbard model on the two-dimensional square lattice is given by
\begin{equation}\label{Hubbardmodel}
H=-t_1\sum^{}_{\langle{i},j\rangle,\sigma}c^{\dag}_{i\sigma}c_{j\sigma}-t_2\sum^{}_{\langle\langle{}i,j\rangle\rangle,\sigma}c^{\dag}_{i\sigma}c_{j\sigma}+U\sum^{}_{i}n_{i\uparrow}n_{i\downarrow},
\end{equation}
where  $c^{\dag}_{i\sigma} (c_{i\sigma}$)  creates (annihilates) an electron at site $i$ with spin $\sigma$, and $n_{i\sigma}=c^{\dag}_{i\sigma}c_{i\sigma}$ is the number operator.  $t_1$ and $t_2$ denote the nearest-neighbour and next-nearest-neighbour hoppings, respectively, and $U$ is the on-site Coulomb repulsion. In this work, $t_1$ is chosen as the energy unit.

We employ the mean field approximation to investigate the one-band Hubbard model (\ref{Hubbardmodel}). At the mean field level, the on-site Coulomb interaction is approximated by
\begin{equation}\label{HFapprox}
	Un_{i\uparrow}n_{i\downarrow}\approx Un_{i\uparrow}\langle{}n_{i\downarrow}\rangle+Un_{i\downarrow}\langle{}n_{i\uparrow}\rangle-U\langle{}n_{i\uparrow}\rangle\langle{}n_{i\downarrow}\rangle.
\end{equation}
In this work, the typical CAF and SAF orders as well as the PCAF, DAF, and PAF orders are taken into account. These spin configurations are displayed in Fig.~\ref{SpinConfig}. Details of the mean field derivations, the unit cells and the coordinates we chose, corresponding magnetic wave vectors for different magnetic states can be found in the Appendix~\ref{HFmethod_appendix}.

\section{results}
\subsection{Phase Diagram}
In order to systematically investigate the influence of frustration on the magnetic properties of electron systems, we have calculated the $U/t_1-t_2/t_1$ phase diagram of the Hubbard model on the square lattice at half filling and zero temperature by comparing total energies of various magnetic orderings we studied, including the nonmagnetic (NM), CAF, SAF, DAF, PCAF and PAF states.  Fig.~\ref{Totalenergies} (a) displays the total energies of these antiferromagnetic states as a function of $t_2$ at $U/t_1=4$, where the energy of the DAF state was set to zero. Since the NM state is much higher in energy than other phases, it is not shown here. As expected, due to weak geometric frustration, the CAF and SAF states are the stablest in the small ($t_2/t_1<0.68$) and large ($t_2/t_1>0.83$) $t_2$ regions, respectively. When $0.68<t_2/t_1<0.83$, the effect of geometric frustration is enhanced due to the growing competition between the CAF and SAF states. And macroscopic degeneracies appears especially around $t_2/t_1=1/\sqrt{2}$. This frustration is believed to dramatically affect the magnetism of the one-band Hubbard model we investigated. As shown in Fig~\ref{Totalenergies} (a), instead of the direct transition from the CAF state to SAF state reported in the early mean field studies~\cite{Yu2010PRB}, two new antiferromagnetic states are found to be stabilized in the intermediate region, namely the DAF state for $0.68<t_2/t_1<0.73$ and the PAF state for $0.73<t_2/t_1<0.83$. From the inset of Fig.~\ref{Totalenergies} (b), the PAF state is found to be the ground state in a rather large region of $U$ when $t_2/t_1=0.8$. Moreover, in addition to the NM, PAF and SAF states, the PCAF state with double periodicity in one direction compared to that of the CAF state is realized with the increase of $U$ at $t_2/t_1=0.8$, as depicted in Fig.~\ref{Totalenergies} (b). The calculated mean field phase diagram was summarized in Fig.~\ref{phase_diagram}.

\begin{figure}[tbhp]
\centering
	\includegraphics[width=1\textwidth]{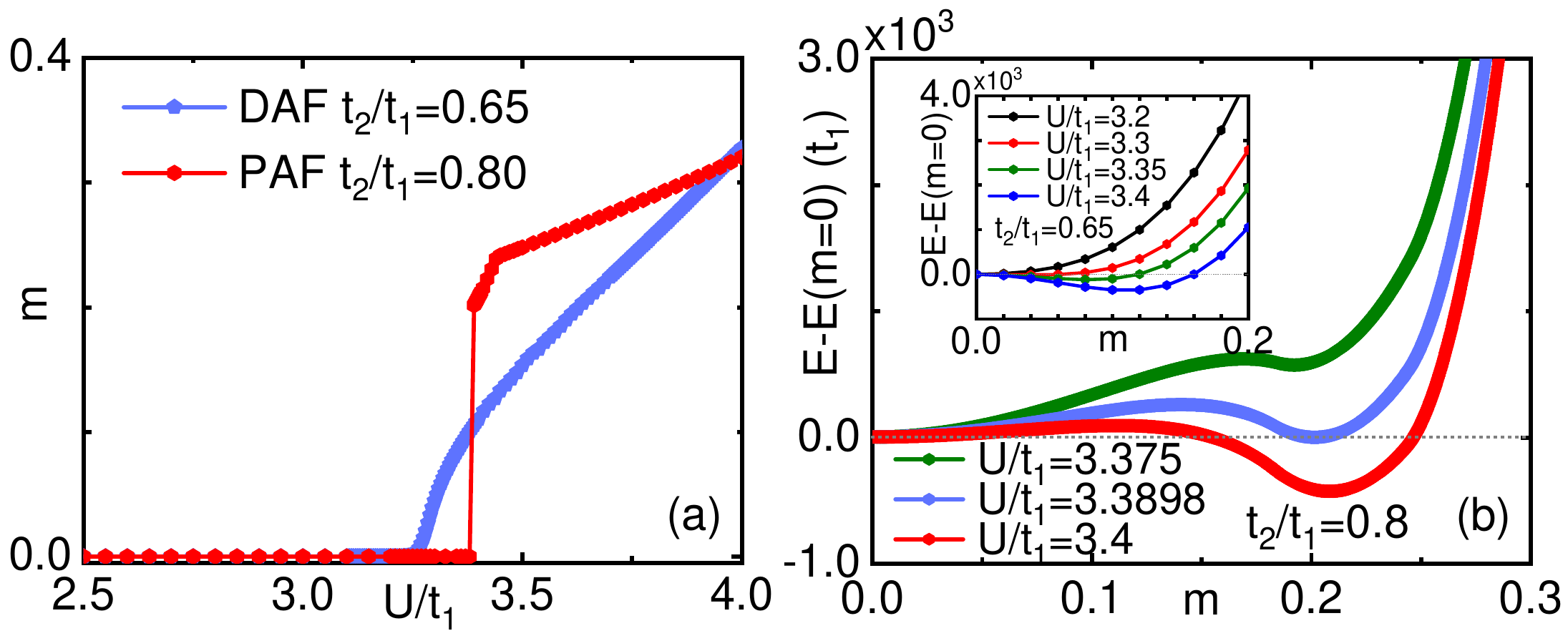}
	\caption{(color online) (a) Magnetic order parameters for the DAF state at $t_2/t_1=0.65$ and the PAF state at $t_2/t_1=0.8$. (b) The free energy of the PAF state as a function of magnetic moment $m$ at $t_2/t_1=0.8$ in the vicinity of the critical point for the phase transition from NM state to the PAF state, where the free energy at $m=0$ is set to zero. The inset denotes the similar quantities for the DAF states at $t_2/t_1=0.65$.}
	\label{moment}
\end{figure}

While common CAF, SAF states are frequently studied in previous theoretical studies~\cite{JPSJ.65.2559,Hofstetter}, the PCAF state was seldom mentioned~\cite{Mizusaki2006PRB,Yamada2013PRB} and the DAF and PAF states have never been explored in the one-band Hubbard model on the frustrated square lattice with $t_1$ and $t_2$. The existence of PCAF order can be understood from the strong coupling limit, where the model (\ref{Hubbardmodel}) can be mapped onto a Heisenberg model with nearest-neighbour $J_1=\frac{4t^2_1}{U}$ and next-nearest-neighbour $J_2=\frac{4t^2_2}{U}$ exchanges. By analysing the energies of the Heisenberg model in the classical limit, the antiferromagnetic orderings with $\mathbf{Q}=(\pi,\pi/n)$, where $n=1,\cdots,\infty$, are exactly degenerate at $J_1=2J_2$ (i.e. $t_2=t_1/\sqrt{2}$)~\cite{Yamada2013PRB}, which are expected to still survive in the critical region of around $t_2=t_1/\sqrt{2}$ even for intermediate $U$. Indeed, the PCAF state is realized between the CAF and SAF states in our calculations, consistent with PIMC results~\cite{Mizusaki2006PRB} and VCA results on certain cluster~\cite{Yamada2013PRB}. The DAF and PAF states can not be obtained from above frustrated spin model. The occurrence of both the DAF and PAF states requires finite third-neighbour spin interaction $J_3$~\cite{Ma2009PRL}, related to the third-neighbour hopping $t_3$ which is absent in the present model, and high-order biquadratic exchanges~\cite{Hu2012PRB,Glasbrenner2015NatPhy} if only spin degrees of freedom are involved. However, situation becomes different when original itinerant model, i.e., the model (\ref{Hubbardmodel}), is taken into account. On one hand, in the large $U$ expansion, higher-order terms can not be neglected at finite $U$, which may act as the long-range and high-order spin exchanges. On the other hand, multiple instabilities appears in the Pauli susceptibility at $U=0$ of the model (\ref{Hubbardmodel}) around the $\mathbf{Q}$ vectors for DAF, PAF and PCAF, indicating strong tendencies towards these symmetry-breaking states, provided incommensurate spin density wave are not considered. Though existence of the DAF, PAF states in the model (\ref{Hubbardmodel}) seems natural, these are completely ignored in the early studies~\cite{Mizusaki2006PRB,Tocchio2008PRB,Yu2010PRB,Nevidomskyy2008PRB,Yoshikawa2009PRB,Yamada2013PRB,Yamaki2013PRB,Misumi2016JPSJ}. Therefore, the nature of the DAF and PAF states is still unknown.
\subsection{The DAF and PAF States}
\begin{figure}[tbhp]
\centering
	\includegraphics[width=0.9\textwidth]{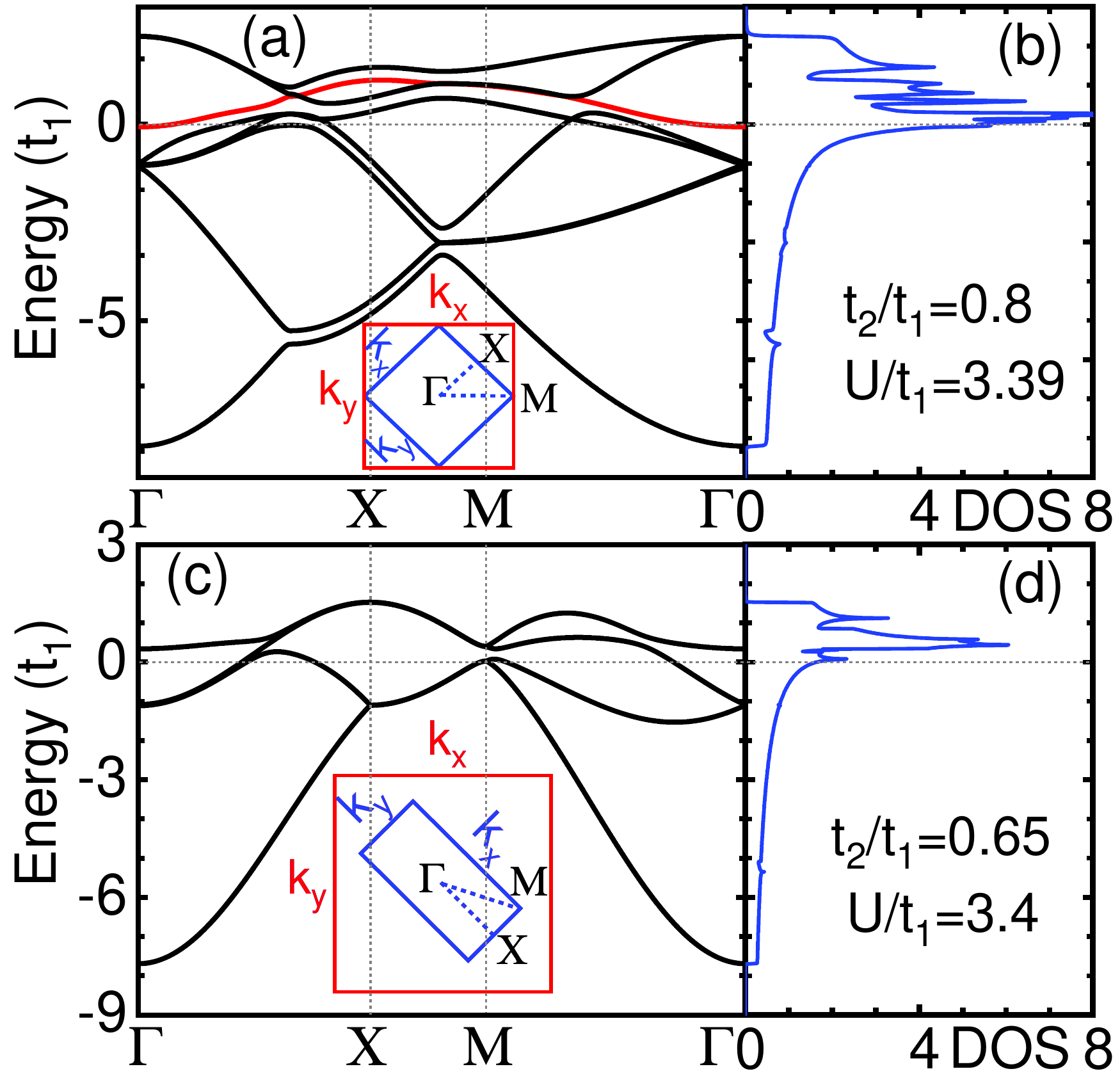}
	\caption{(color online) Band structures and corresponding density of states (DOS) for the PAF state at $t_2/t_1=0.8$ and $U/t_1=3.39$ (a)-(b) and for the DAF state at $t_2/t_1=0.65$ and $U/t_1=3.4$ (c)-(d). The high-symmetry path in the Brillouin zone used to calculate the band structure is shown in the insets of (a) for the PAF state and (c) for the DAF state, respectively.  }
	\label{banddos}
\end{figure}
Now, we focus on the properties of the DAF and PAF states. Fig.~\ref{moment} (a) displays the magnetic order parameters of the DAF and PAF states at $t_2/t_1=0.65$ and $t_2/t_1=0.8$, respectively. The magnetic moment exhibits distinct behaviors in these two phases as a function of $U$. For the case of the DAF state, the order parameter continuously increases with the increase of $U/t_1$. In contrast, there is a sudden jump in the magnetic moment at around $U_{\rm c}(t_2/t_1=0.8)\approx3.39t_1$ for the PAF state. This dramatic difference is attributed to the distinct behaviors of total energies as a function of magnetic moment. As shown in Fig.~\ref{moment} (b), when $t_2/t_1=0.8$, for various $U/t_1$ the total energies of the PAF state always have two local minimums, separating by a finite potential barrier, where one minimum is at $m=0$ and the another is at finite $m$. The fact that the minimum of finite magnetic moment becomes a global minimum when crossing the NM-PAF transition suggests a first-order phase transition. In contrast, at $t_2/t_1=0.65$, the total energies of the DAF state have only one local minimum for all $U/t_1$, as displayed in the inset of Fig.~\ref{moment} (b). Obviously, the magnetic moment continuously vanishes at the transition, indicating the NM-DAF phase transition is of second order. The nature of these transitions is further confirmed by calculating the derivative of the total energy with respect to the Coulomb interaction $U$ (not shown). Similarly, the nature of the phase transitions presented in this work are determined. As shown in Fig.~\ref{phase_diagram}, the white and black lines denote the first- and second-order phase transitions, respectively. Furthermore, as displayed in Fig.~\ref{moment} (a), a kink appears in the magnetic moment for the case of the PAF state, which is attributed to the Lifshits transition induced by a Van Hove singularity lifted above the Fermi level, as discussed below.

\begin{figure}[tbhp]
\centering
	\includegraphics[width=0.7\textwidth]{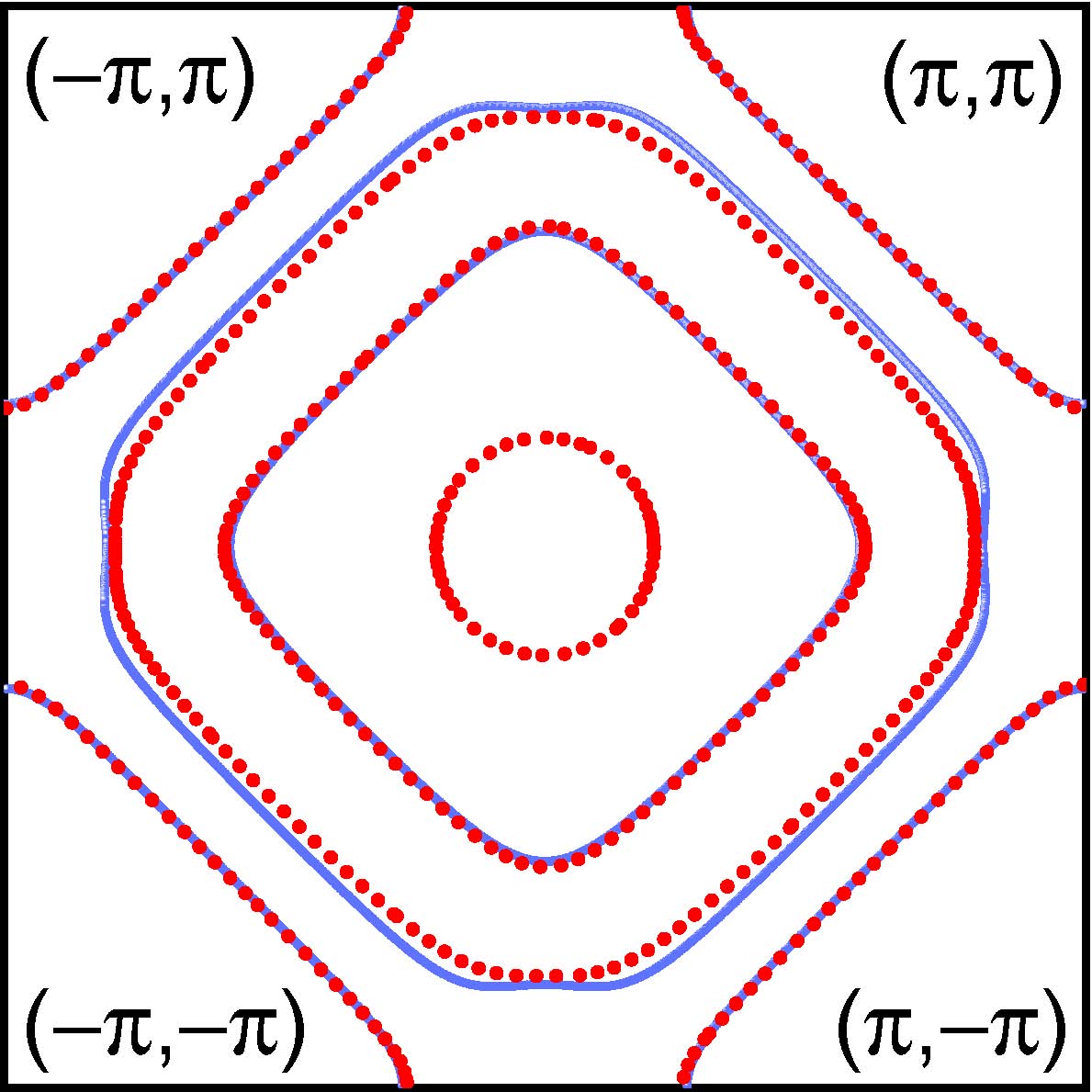}
	\caption{(color online) Fermi surfaces of the PAF order at $t_2/t_1=0.8$ for $U/t_1=3.4$ (red dashed lines) and $3.5$ (blue solid lines). The change of the Fermi surface topology  suggests the occurrence of the Lifshits transition.}
	\label{Fermisurface}
\end{figure}

In order to gain a deep insight into the properties of the DAF and the PAF states, we have calculated the band structures and the density of states (DOS) for the PAF state at $t_2/t_1=0.8$ and $U/t_1=3.39$ and for the DAF state at $t_2/t_1=0.65$ and $U/t_1=3.4$, as displayed in Fig.~\ref{banddos}. Both band structure and DOS suggest that the system is metallic in both PAF and DAF states. Particularly, as shown in Fig.~\ref{banddos} (a), an electron pocket is present near the $\Gamma$ point in the Brillouin zone (see the red band around the Fermi level). As the electronic repulsion further increases, this band is lifted above the Fermi level and corresponding pocket completely vanishes, leading to the occurrence of the Lifshits transition, i.e., disappearance of inner Fermi surface in the vicinity of the $\Gamma$ point, which is shown in Fig.~\ref{Fermisurface}. The Van Hove singularity (Fig.~\ref{banddos} (b)) lifted above the Fermi level gives rise to the appearance of the Lifshits transition, resulting in the kink of the magnetic moment for the PAF state at $t_2/t_1=0.8$ (Fig.~\ref{moment} (a)). Similar results were observed in the CAF state, consistent with existing mean field results~\cite{Yu2010PRB}. However, the Lifshits transition does not take place in the DAF state at $t_2/t_1=0.65$.

\section{Discussions}
\begin{figure}[htbp]
\centering
	\includegraphics[width=0.7\textwidth]{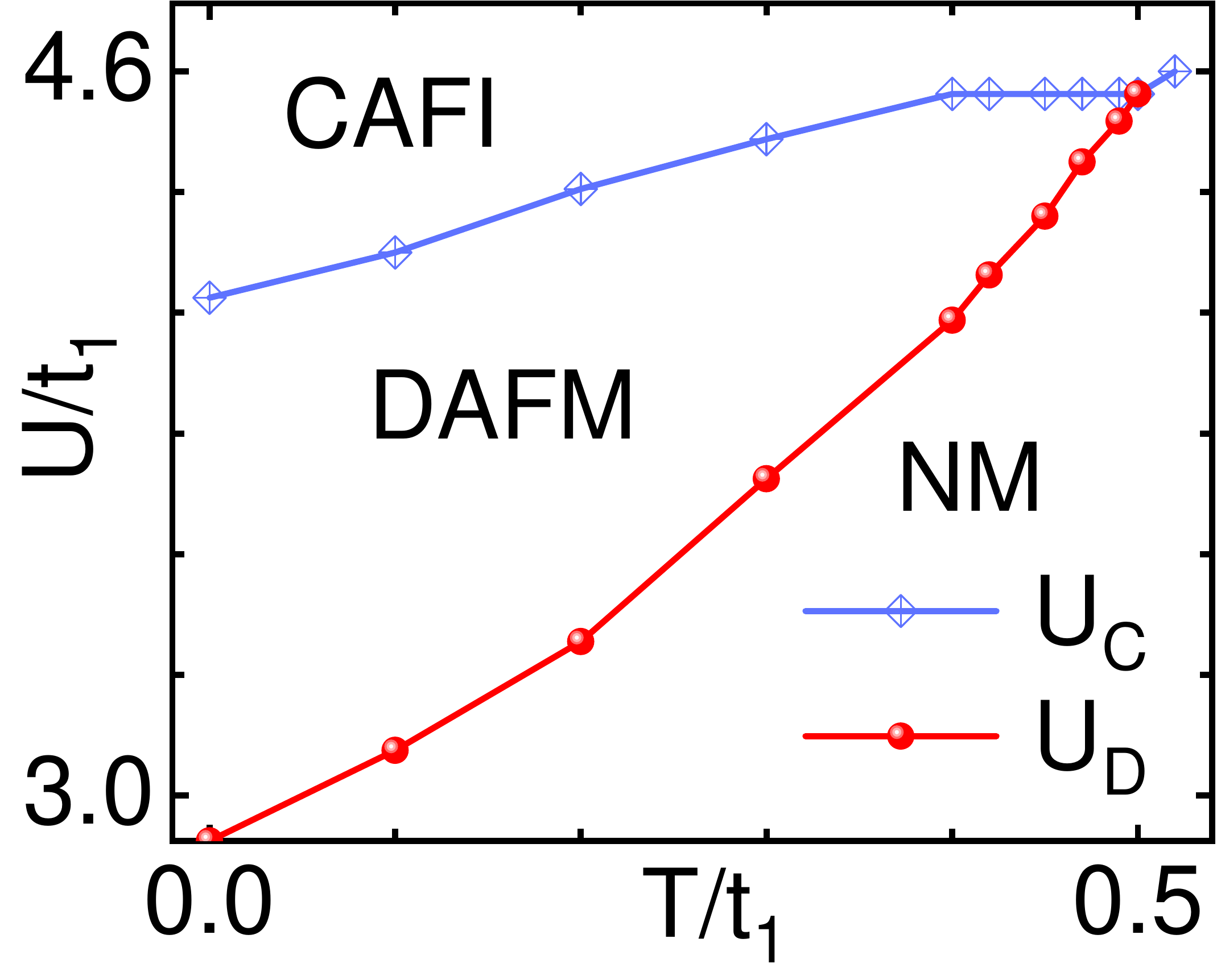}
	\caption{(color online) $U/t_1$-$T/t_1$ phase diagram at $t_2/t_1=1/\sqrt{2}$, where $U_{\mathrm{C}}$ and $U_{\mathrm{D}}$ denote the critical points for the appearance of the CAF and DAF states, respectively.}
	\label{ThermalStability}
\end{figure}
In this work, the influence of geometric frustration on the magnetic properties of the Hubbard model on the square lattice with $t_1$ and $t_2$ is systematically investigated by the mean field approximation at half filling and zero temperature. We focus on the parameter space of the intermediate electronic repulsions. The DAF and PAF states are found to be stabilized at small $U$ in the vicinity of $t_2=t_1/\sqrt{2}$. This is in sharp contrast to the slave-boson mean field results~\cite{Li2012JMMM}, where the third-neighbour hopping $t_3$, not considered in the present work, is found to  be indispensable for the appearance of the DAF state. In spite of $t_3$ being absent, the energetic stability of the DAF and PAF phases against other antiferromagnetic states is further confirmed at finite temperature $T$. The result for the DAF state at $t_2=t_1/\sqrt{2}$ is shown in Fig.~\ref{ThermalStability}. It is found that the DAF state is sandwiched between the CAF insulating state and the NM phase at low temperature. As $T$ increases, the region for the DAF state is shrunk and vanishes at the triple point $T$ of about 0.5$t_1$. Similar temperature-dependent behavior was also found for PAF ordered state. These results suggest that the DAF and PAF state may be realized in real materials with strong geometric frustration and moderate electronic interaction.

\begin{figure}[htbp]
\centering
	\includegraphics[width=0.7\textwidth]{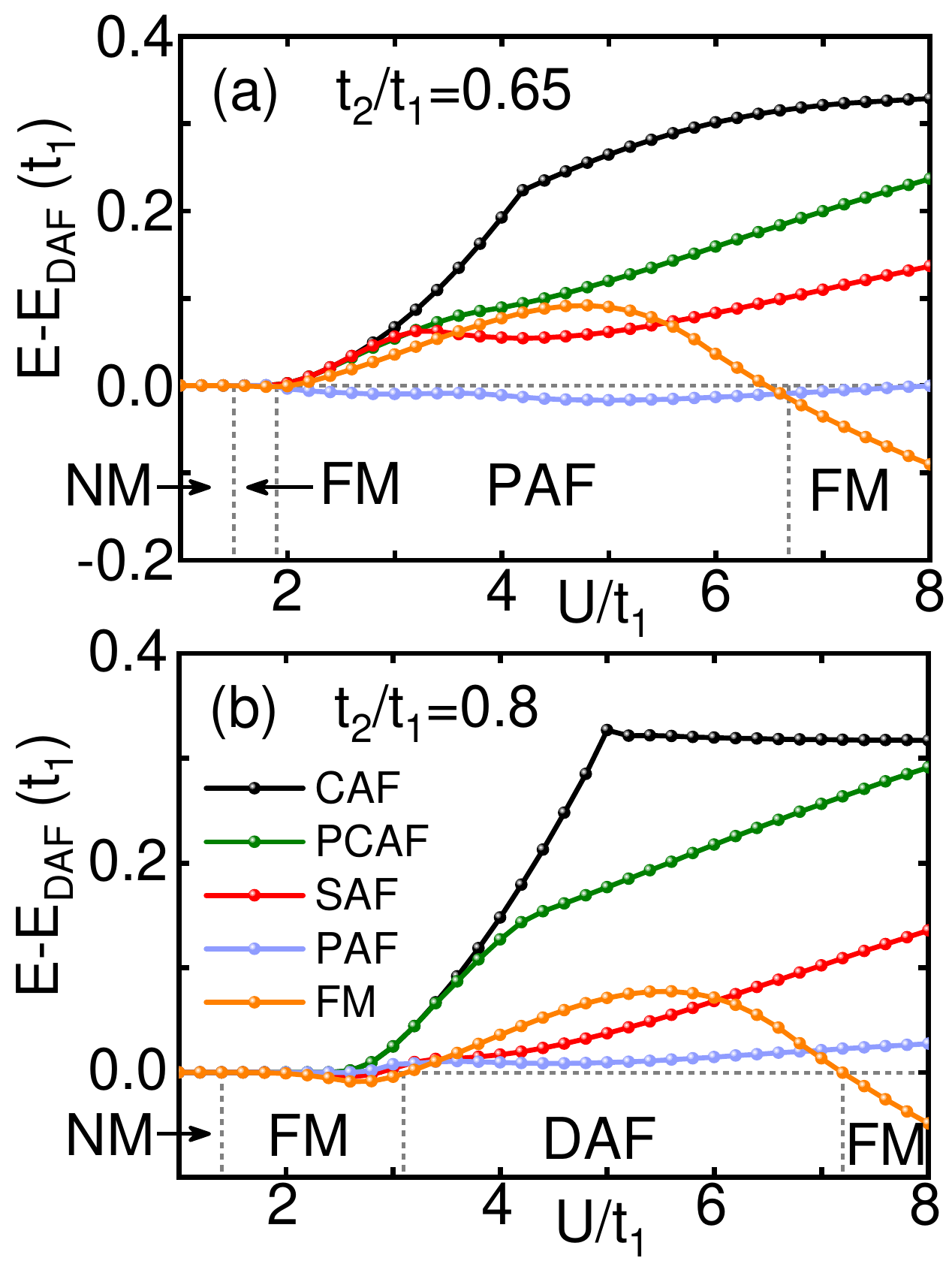}
	\caption{(color online) The total energies of various magnetic states including the CAF, PCAF, SAF, PAF and ferromagnetic (FM) states at filling $n=1.2$ for $t_2/t_1=0.65$ (a) and $t_2/t_1=0.8$ (b). At small $U$, the magnetic moment vanishes for all magnetic orderings and the system is in the NM state.}
	\label{doping_effect}
\end{figure}

Except for the DAF, PAF and PCAF phases, the calculated phase digram shown in Fig.~\ref{phase_diagram} is quantitatively consistent with previous mean field results~\cite{Yu2010PRB} and in qualitative agreement with that obtained by other powerful methods, like PIMC~\cite{Mizusaki2006PRB}, VMC~\cite{Tocchio2008PRB}, VCA~\cite{Yamada2013PRB}, suggesting that the mean field approximation produces reliable results as the quantum fluctuations is substantially reduced in broken-symmetry systems~\cite{Zhang2012PRB}. Moreover, the PCAF state is realized between the CAF and SAF states, qualitatively coinciding with the PIMC results~\cite{Mizusaki2006PRB} at large $U$ and the VCA results~\cite{Yamada2013PRB} on $3\times4$ cluster at intermediate $U$, which further justifies the reliability of the mean field approximation. However, since the quantum fluctuations are completely frozen in the mean field approximation, this method can not realize the paramagnetic Mott insulator observed by PIMC~\cite{Mizusaki2006PRB} around the parameter space for the DAF and PAF states. However, in this region, such a spin disordered state was also not reproduced by the VMC~\cite{Tocchio2008PRB} and VCA~\cite{Yamada2013PRB} methods. It is covered by NM, CAF, SAF, and possible PCAF states and the paramagnetic Mott insulator only appears at large onsite Coulomb repulsion. Since the DAF and PAF states have never been taken into account in previous studies~\cite{Mizusaki2006PRB,Tocchio2008PRB,Yu2010PRB,Nevidomskyy2008PRB,Yoshikawa2009PRB,Yamada2013PRB,Yamaki2013PRB,Misumi2016JPSJ} and these states are the ground states with total energies well separated from other states in the respective region of the phase diagram within mean field approximation, we argue that the DAF and PAF states may also win the competition among all the states we considered even beyond mean field approximation.

Finally, we discuss the relevance of the phase diagram, displayed in Fig.~\ref{phase_diagram}, to real materials. It is well known that the one-band Hubbard model (\ref{Hubbardmodel}) is the basic model to describe the magnetism in cuprates. Owing to the fact that $t_2/t_1$ is small in cuprates, the CAF insulating state successfully accounts for the antiferromagnetic order observed in these compounds. Concerning the iron-based superconductors, the onsite Coulomb interactions are believed to be moderate~\cite{Qazilbash2009NatPhy,Yin2011NatMat} and the geometric frustration $t_2/t_1$ of different orbitals varies from $\sim0$ to $>1$ as reported in an {\it ab} initio study~\cite{JPSJ2010Miyake}, which may support possible tendency towards all types of magnetic orders presented in our phase diagram, for example the DAF and PAF order as observed in FeTe~\cite{Li2009PRB,Ma2009PRL,Tam2019PRB,Ducatman2012PRL}, the PCAF order in FeSe~\cite{Cao2015PRB}, and SAF order in many others~\cite{nature2008Cruz,Huang2008PRL,PRB2009Li}. Then, the total magnetic state of the whole system may result from interplay among the tendencies towards different magnetic states in different orbitals, reminiscent of weak antiferromagnetism induced by coupling of frustrated and unfrustrated bands~\cite{Lee2010PRB}.

The reason why a one-band model can be applied to qualitatively understand the physical properties of a multi-orbital system is the following. Since the Hund's coupling which serves as a band decoupler suppresses the orbital fluctuations~\cite{Medici2011PRB,Song2017PRB}, the multi-orbital system can be viewed as a collection of single-band ones, leading to a unified phase diagram shared by iron-based superconductors and cuprates~\cite{Medici2014PRL}. Vice verse, the one-band Hubbard model~(\ref{Hubbardmodel}) can be mapped onto an effective multi-orbital problem, leading to a unified understanding of non-Fermi liquid behaviors in both high-T$_c$ superconducting families~\cite{Werner2016}. These results, together with ours, indicate that the one-band Hubbard model on frustrated square lattice with hoppings up to next-nearest neighbour may be the minimum model to describe the magnetism of both iron-based superconductors and cuprates.

Given the fact that parent compounds of iron-based superconductors are electron-doped systems due to $5$ Fe 3$d$ orbitals filled by six electrons, we have investigated the effect of electron doping on the DAF and PAF states by calculating the energy differences of the magnetic orderings considered in the work at filling $n=1.2$ for $t_2/t_1=0.65$ and $t_2/t_1=0.8$. Additionally, the ferromagnetic (FM) state is also taken into account, which is found to be stabilized at weak and intermediate $U$. The result is shown in Fig.~\ref{doping_effect}, where the energy of the DAF state is set to zero. It is found that both the DAF and PAF states are favored by doping since the regions for these two phases are remarkably enlarged in comparison to that at half filling. It should be noted that, when $t_2/t_1=0.65$, the PAF state rather than the DAF state is realized at filling $n=1.2$ in contrast to the case of half filling. Similar conclusions are obtained for $t_2/t_1=0.8$.

Last point we have to emphasize is that the AF states we studied in this paper are restricted to CAF, SAF, DAF, PCAF, and PAF states, which are observed experimentally in mother compounds of high-T$_c$ cuprates and iron-based superconductors.

\section{conclusion}
In conclusion, we have studied the two-dimensional Hubbard model on the square lattice with $t_1$ and $t_2$ at half filling and zero temperature within the mean field theory. We find that the DAF and PAF states are realized at moderate electronic repulsion in the vicinity of $t_2/t_1=1/\sqrt{2}$ as a result of the interplay between geometrical frustration and kinetic energy, electronic Coulomb interaction. In the DAF and PAF states, the system is a metal. Moreover, the DAF and PAF states are robust under finite temperature and finite doping. Since the CAF and DAF, PAF, PCAF, SAF states we obtained can reasonably account for the corresponding magnetic orderings in cuprates and parent compounds of iron-based superconductors, respectively, we argue that the Hubbard model (\ref{Hubbardmodel}) may be the minimum model which can capture the underlying physics of both high-T$_c$ superconductors. Our findings call for further work on the new phase diagram beyond mean field approximation.

\begin{ack}
This work is financially supported by the National Natural Science Foundation of China (Grant No. 11774258, 12004283) and Postgraduate Education Reform Project of Tongji University (Grant No. GH1905), Z. Y. Song acknowledges the financial support by China Postdoctoral Science Foundation (Grant No. 2019M651563).
\end{ack}

\appendix
\begin{figure}[htbp]
\centering
	\includegraphics[width=0.9\textwidth]{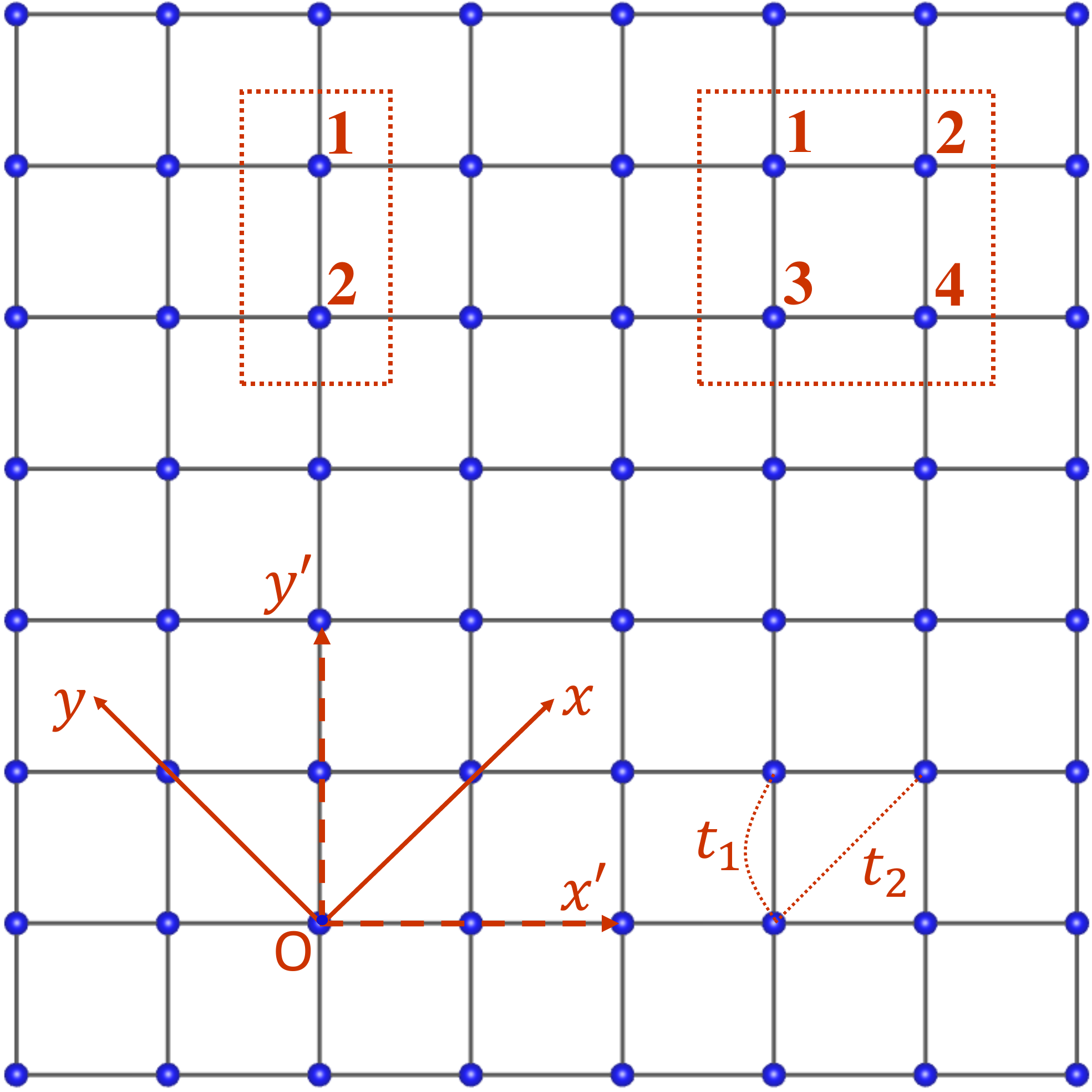}
	\caption{(color online) Cartoon for the square lattice showing the unit cells and the coordinates we used for different magnetic states. The unit cell with two sites is for CAF, DAF, SAF, and PCAF states, while that with four sites for PAF state. The coordinates with $x$ and $y$ axes along diagonal directions of the quare lattice is used for CAF, DAF, and SAF states, while along bond directions for PCAF and PAF ones. $t_1$ and $t_2$ represent for nearest and next-nearest neighbour hoppings, respectively.}
	\label{sqlattice}
\end{figure}
\section{The mean field approximation to the Hubbard model on frustrated square lattice}\label{HFmethod_appendix}
The two-dimensional Hubbard model with nearest-neighbour $t_1$ and next-nearest-neighbour $t_2$ hoppings on the square lattice, which is defined by Eq (\ref{Hubbardmodel}), is studied within the mean field approximation, where the onsite Coulomb repulsion is approximately treated via Eq (\ref{HFapprox}).
Without considering charge ordered state, the average occupation per site is expressed as
\begin{equation}\label{occupation}
	\bar{n}=\langle{}n_{li\uparrow}\rangle+\langle{}n_{li\downarrow}\rangle,
\end{equation}
where  $l$ represents the $l$-th unit cell, and $i=1,\cdots,n_{\mathrm{c}}$ the sublattice index with $n_{\mathrm{c}}$ the total number of lattice sites in a unit cell.

The magnetic order parameter $m_i$ of each sublattice is defined as
\begin{equation}\label{magwavevector}
m_ie^{i\mathbf{Q}\cdot\mathbf{l}}=\langle{}n_{li\uparrow}\rangle-\langle{}n_{li\downarrow}\rangle,
\end{equation}
where $\mathbf{Q}$ denotes the magnetic wave vector. If $m=0$, the system is in the NM state.

Combing Eq (\ref{magwavevector}) and Eq (\ref{occupation}), we have
\begin{equation}
	\langle{}n_{li\uparrow}\rangle=\frac{\bar{n}}{2}+\frac{m_i}{2}e^{i\mathbf{Q}\cdot\mathbf{l}} ~\mathrm{and}~
	\langle{}n_{li\downarrow}\rangle=\frac{\bar{n}}{2}-\frac{m_i}{2}e^{i\mathbf{Q}\cdot\mathbf{l}},
\end{equation}
Finally, at the mean field level, the interaction part of the Hamiltonian (\ref{Hubbardmodel}) can be written as
\begin{eqnarray}
	H_{\mathrm{int}}=&U\sum_{li}\left(\frac{\bar{n}}{2}-\frac{m_i}{2}e^{i\mathbf{Q}\cdot\mathbf{l}}\right)n_{li\uparrow}	+\nonumber\\
	&U\sum_{li}\left(\frac{\bar{n}}{2}+\frac{m_i}{2}e^{i\mathbf{Q}\cdot\mathbf{l}}\right)n_{li\downarrow}+\nonumber\\
	&U\sum_{li}\frac{m^2_i}{4}e^{i2\mathbf{Q}\cdot\mathbf{l}}-U\frac{Nn_c\bar{n}^2}{4},
\end{eqnarray}
where $N$ denotes the total number of unit cells. For all the antiferromagnetic orderings considered in the work, $e^{i2\mathbf{Q}\cdot\mathbf{l}}=1$.  After performing the Fourier transformation, we obtain
\begin{eqnarray}\label{EHartree}
H_{\mathrm{int}}=&U\sum_{ki}\left(\frac{\bar{n}}{2}n_{ki\uparrow}-\frac{m_i}{2}c^{\dag}_{ki\uparrow}c_{k+\mathbf{Q}i\uparrow}\right)	+\nonumber\\
&U\sum_{ki}\left(\frac{\bar{n}}{2}n_{ki\downarrow}+\frac{m_i}{2}c^{\dag}_{ki\downarrow}c_{k+\mathbf{Q}i\downarrow}\right)+\nonumber\\
&NU\sum_{i}\frac{m^2_i}{4}-U\frac{Nn_c\bar{n}^2}{4}.
\end{eqnarray}

It is expected that the symmetry-broken magnetic orderings occur when the electronic interaction exceeds a threshold. In this work, paramagnetic and five antiferromagnetic states are considered, including the CAF, DAF, SAF, PCAF and PAF phases. The unit cell consisting of two lattice sites is chosen to do the mean field calculations for all antiferromagnetic states except for the PAF state, for which the corresponding calculations are done with unit cell containing four lattice sites. These two kinds of unit cells are displayed in Fig.~\ref{sqlattice}. For the CAF, DAF and SAF states, the $x(y)$ is along the square diagonal direction, i.e. ox (oy) direction. For the PCAF and PAF states,  the $x(y)$ is along the nearest-neighbour bond direction, i.e. ox$^\prime$ (oy$^\prime$) direction. The corresponding $\mathbf{Q}$ vectors of CAF, DAF, SAF, PCAF, and PAF states are $(0,0)$, $(\pi,0)$, $(\pi,\pi)$, $(\pi,\pi)$, and $(\pi,\pi)$, respectively.

When the unit cell has two lattice sites, the noninteracting part of the Hamiltonian can be written as
\begin{equation}
H_0=\sum_{k\sigma}(c^{\dag}_{k1\sigma},c^{\dag}_{k2\sigma})
\left(
\begin{array}{ccc}
\epsilon_{11}(k)&\epsilon_{12}(k)\\\epsilon_{21}(k)&\epsilon_{22}(k)
\end{array}
\right)
\left(
\begin{array}
c_{k1\sigma}\\c_{k2\sigma}
\end{array}
\right),
\end{equation}
where $\sigma$ denotes the electron spin. For the CAF, DAF and SAF states,
\begin{eqnarray}
\epsilon_{11}(k)&=-2t_2(\cos{}k_x+\cos{}k_y)=\epsilon_{22}(k),\\
\epsilon_{12}(k)&=-4t_1e^{i\frac{k_x+k_y}{2}}\cos\frac{k_x}{2}\cos\frac{k_y}{2}=\epsilon^*_{21}(k).
\end{eqnarray}
For the PCAF state,
\begin{eqnarray}
	\epsilon_{11}(k)&=-2t_1\cos{}k_x=\epsilon_{22}(k),\\
	\epsilon_{12}(k)&=-t_1(1+e^{-ik_y})-2t_2\cos{k_x}(1+e^{-ik_y}),
\end{eqnarray}
and $\epsilon^*_{21}(k)=\epsilon_{12}(k)$. For the case of the unit cell containing four lattice sites, the kinetic energy can be written in a matrix form

\begin{equation}
H_0=
\sum_{k\sigma}\langle\Psi_{k\sigma}|
\left(
\begin{array}{ccc}
\epsilon_{11}(k)&\cdots&\epsilon_{14}(k)\\\vdots&&\vdots\\\epsilon_{41}(k)&\cdots&\epsilon_{44}(k)
\end{array}
\right)
|\Psi_{k\sigma}\rangle,
\end{equation}

where $|\Psi_{k\sigma}\rangle^{T}=(c_{k1\sigma},c_{k2\sigma},c_{k3\sigma},c_{k4\sigma})$, $\epsilon_{11}(k)=0$, $\epsilon_{12}(k)=-t_1(1+e^{ik_x})$, $\epsilon_{13}(k)=-t_1(1+e^{-ik_y})$, $\epsilon_{14}(k)=-t_2(1+e^{ik_x}+e^{-ik_y}+e^{i(k_x-k_y)})$, $\epsilon_{22}(k)=0$, $\epsilon_{23}(k)=-t_2(1+e^{-ik_x}+e^{-ik_y}+e^{-i(k_x+k_y)})$, $\epsilon_{24}(k)=-t_1(1+e^{-ik_y})$, $\epsilon_{33}(k)=0$, $\epsilon_{34}(k)=-t_1(1+e^{ik_x})$, $\epsilon_{44}(k)=0$. And $\epsilon_{ji}(k)=\epsilon^*_{ij}(k)$ with $j>i$. In general, the whole Hamiltonian is
\begin{equation}\label{SCFone}
H=H_{\mathrm{int}}+H_0,
\end{equation}
in the basis $(|\Psi_{k\sigma}\rangle,|\Psi_{k+\mathbf{Q}\sigma}\rangle)$. Due to the occurrence of band folded when $\mathbf{Q}\neq(0,0)$, $k$ will be limited to the corresponding reduced Brillouin zone (RBZ).

In the mean field calculations, the key quantities are the average occupation $\bar{n}$ and magnetic order parameter $m_i$. They can be self-consistently obtained by diagonalizing the total Hamiltonian with the unitary transformation $S$, which is defined via $c^{\dag}_{ki\sigma}=\sum_{\alpha}S_{ki\alpha}\phi^{\dag}_{k\alpha\sigma}$ where $\phi^{\dag}_{k\alpha\sigma}$ create an electron with spin $\sigma$ in band $\alpha$ at momentum $k$. In the band basis, the ground state can be expressed as
\begin{equation}
|\rangle=\prod_{\alpha\sigma{}k\in\mathrm{RBZ},E_{k\alpha\sigma}\leq E_{\mathrm{f}}}\phi^{\dag}_{k\alpha\sigma}|\mathrm{vac}\rangle,	
\end{equation}
where $E_{k\alpha\sigma}$ and $E_{\mathrm{f}}$ are the band eigen energy and the Fermi energy, respectively. The total energy per magnetic unit cell is
\begin{eqnarray}\label{Etotal}
E_{\mathrm{total}}=&\frac{1}{N}\sum_{\alpha\sigma{}k\in{\mathrm{RBZ}}}E_{k\alpha\sigma}\theta(E_{\mathrm{f}}-E_{k\alpha\sigma}) \nonumber\\
&+U\sum_{i}\frac{m^2_i}{4}-\frac{Un_c\bar{n}^2}{4},
\end{eqnarray}
where $\theta(E_{\mathrm{f}}-E_{k\alpha\sigma})$ is the step function. The average occupation is
\begin{eqnarray}
\bar{n}&=\frac{1}{Nn_{\mathrm{c}}}\sum_{li\sigma}\langle|{n_{li\sigma}}|\rangle \nonumber\\
&=\frac{1}{Nn_{\mathrm{c}}}\sum_{i\sigma{}k\in{\mathrm{RBZ}}}(\langle|{n_{ki\sigma}}|\rangle+\langle|{n_{k+\mathbf{Q}i\sigma}}|\rangle).
\end{eqnarray}
It is obvious that
\begin{eqnarray}\label{SCFtwo}
\bar{n}=\frac{1}{Nn_{\mathrm{c}}}\sum_{\alpha\sigma{}k\in{\mathrm{RBZ}}}\theta(E_{\mathrm{f}}-E_{k\alpha\sigma}).
\end{eqnarray}

Similarly, the magnetic order parameter is
\begin{eqnarray}
m_i=\frac{1}{N}\sum_{k}\langle|c^{\dag}_{ki\uparrow}c_{k+\mathbf{Q}i\uparrow}|\rangle-\frac{1}{N}\sum_{k}\langle|c^{\dag}_{ki\downarrow}c_{k+\mathbf{Q}i\downarrow}|\rangle.
\end{eqnarray}
By using $c^{\dag}_{ki\sigma}=\sum_{\alpha}S_{ki\alpha}\phi^{\dag}_{k\alpha\sigma}$, we have
\begin{eqnarray}\label{SCFthree}
m_i=&\frac{2}{N}\sum_{\alpha{}k\in{\mathrm{RBZ}}}\{Re[S_{ki\alpha\uparrow}S^{*}_{k+Qi\alpha\uparrow}-S_{ki\alpha\downarrow}S^{*}_{k+Qi\alpha\downarrow}]\}\nonumber\\
&\times\theta(E_{\mathrm{f}}-E_{k\alpha}).
\end{eqnarray}
Note $E_{k\alpha}=E_{k\alpha\uparrow}=E_{k\alpha\downarrow}$ for all antiferromagnetic states and $Re[ ]$ means taking the real part. Obviously, steady solutions of different magnetic states can be obtained by self-consistently solving Eq. (\ref{SCFone}), (\ref{SCFtwo}) and (\ref{SCFthree}), and the ground state is determined by comparing Eq. (\ref{Etotal}) averaged by total number of lattice sites per unit cell among all the magnetic sates we considered.

At finite temperature, the free energy can be expressed as

\begin{equation}\label{Freeenergy}
F=-\frac{1}{\beta}ln\Xi+Nn_c\bar{n}\mu+NU\sum_{i}\frac{m^2_i}{4}-U\frac{Nn_c\bar{n}^2}{4},
\end{equation}
where the grand partition function reads

\begin{eqnarray}\label{partition}
ln\Xi=\sum_{\alpha{}k\in{\mathrm{RBZ}\sigma}}ln[1+e^{-\beta(E_{k\alpha\sigma}-\mu)}],
\end{eqnarray}
and $\beta$ is the inversed temperature defined as $1/kT$.
\bibliographystyle{unsrt}
\bibliography{references}

\begin{thebibliography}{10}

\bibitem{Keimer2015}
B.~Keimer, S.~A. Kivelson, M.~R. Norman, S.~Uchida, and J.~Zaanen.
\newblock {From quantum matter to high-temperature superconductivity in copper
  oxides}.
\newblock {\em {Nature}}, {518}({7538}):{179}, {FEB 12} {2015}.

\bibitem{Dai2015RMP}
Pengcheng Dai.
\newblock {Antiferromagnetic order and spin dynamics in iron-based
  superconductors}.
\newblock {\em Rev. Mod. Phys.}, 87:855, Aug 2015.

\bibitem{Lee2006RMP}
Patrick~A. Lee, Naoto Nagaosa, and Xiao-Gang Wen.
\newblock {Doping a Mott insulator: Physics of high-temperature
  superconductivity}.
\newblock {\em Rev. Mod. Phys.}, 78:17, Jan 2006.

\bibitem{Scalapino2012RMP}
D.~J. Scalapino.
\newblock {A common thread: The pairing interaction for unconventional
  superconductors}.
\newblock {\em Rev. Mod. Phys.}, 84:1383--1417, Oct 2012.

\bibitem{GeorgesARCMP2013}
Antoine Georges, Luca~de' Medici, and Jernej Mravlje.
\newblock {Strong Correlations from Hund's Coupling}.
\newblock {\em Annu. Rev. Condens. Matter Phys.}, 4(1):137, 2013.

\bibitem{Medici2011PRB}
Luca de' Medici.
\newblock {Hund's coupling and its key role in tuning multiorbital
  correlations}.
\newblock {\em Phys. Rev. B}, 83:205112, May 2011.

\bibitem{Song2017PRB}
Ze-Yi Song, Xiu-Cai Jiang, Hai-Qing Lin, and Yu-Zhong Zhang.
\newblock {Distinct nature of orbital-selective Mott phases dominated by
  low-energy local spin fluctuations}.
\newblock {\em Phys. Rev. B}, 96:235119, Dec 2017.

\bibitem{Medici2014PRL}
Luca de' Medici, Gianluca Giovannetti, and Massimo Capone.
\newblock {Selective Mott Physics as a Key to Iron Superconductors}.
\newblock {\em Phys. Rev. Lett.}, 112:177001, Apr 2014.

\bibitem{Yu2013PRL}
Rong Yu and Qimiao Si.
\newblock {Orbital-Selective Mott Phase in Multiorbital Models for Alkaline
  Iron Selenides
  ${\mathbf{K}}_{1\ensuremath{-}x}{\mathrm{Fe}}_{2\ensuremath{-}y}{\mathrm{Se}}_{2}$}.
\newblock {\em Phys. Rev. Lett.}, 110:146402, Apr 2013.

\bibitem{Li2009PRB}
Shiliang Li, Clarina de~la Cruz, Q.~Huang, Y.~Chen, J.~W. Lynn, Jiangping Hu,
  Yi-Lin Huang, Fong-Chi Hsu, Kuo-Wei Yeh, Maw-Kuen Wu, and Pengcheng Dai.
\newblock {First-order magnetic and structural phase transitions in
  ${\rm{Fe}}_{1+y}{\rm{Se}}_{x}{\rm{Te}}_{1\ensuremath{-}x}$}.
\newblock {\em Phys. Rev. B}, 79:054503, Feb 2009.

\bibitem{Ma2009PRL}
Fengjie Ma, Wei Ji, Jiangping Hu, Zhong-Yi Lu, and Tao Xiang.
\newblock {First-Principles Calculations of the Electronic Structure of
  Tetragonal $\ensuremath{\alpha}$-FeTe and $\ensuremath{\alpha}$-FeSe
  Crystals: Evidence for a Bicollinear Antiferromagnetic Order}.
\newblock {\em Phys. Rev. Lett.}, 102:177003, Apr 2009.

\bibitem{Tam2019PRB}
David~W. Tam, Hsin-Hua Lai, Jin Hu, Xingye Lu, H.~C. Walker, D.~L. Abernathy,
  J.~L. Niedziela, Tobias Weber, M.~Enderle, Yixi Su, Z.~Q. Mao, Qimiao Si, and
  Pengcheng Dai.
\newblock {Plaquette instability competing with bicollinear ground state in
  detwinned FeTe}.
\newblock {\em Phys. Rev. B}, 100:054405, Aug 2019.

\bibitem{Ducatman2012PRL}
Samuel Ducatman, Natalia~B. Perkins, and Andrey Chubukov.
\newblock {Magnetism in Parent Iron Chalcogenides: Quantum Fluctuations Select
  Plaquette Order}.
\newblock {\em Phys. Rev. Lett.}, 109:157206, Oct 2012.

\bibitem{Zhou2018PRL}
Y.~Zhou, L.~Miao, P.~Wang, F.~F. Zhu, W.~X. Jiang, S.~W. Jiang, Y.~Zhang,
  B.~Lei, X.~H. Chen, H.~F. Ding, Hao Zheng, W.~T. Zhang, Jin-feng Jia, Dong
  Qian, and D.~Wu.
\newblock {Antiferromagnetic Order in Epitaxial FeSe Films on
  ${\mathrm{SrTiO}}_{3}$}.
\newblock {\em Phys. Rev. Lett.}, 120:097001, Feb 2018.

\bibitem{Cao2015PRB}
Hai-Yuan Cao, Shiyou Chen, Hongjun Xiang, and Xin-Gao Gong.
\newblock {Antiferromagnetic ground state with pair-checkerboard order in
  FeSe}.
\newblock {\em Phys. Rev. B}, 91:020504, Jan 2015.

\bibitem{Taylor2013PRB}
A.~E. Taylor, S.~J. Sedlmaier, S.~J. Cassidy, E.~A. Goremychkin, R.~A. Ewings,
  T.~G. Perring, S.~J. Clarke, and A.~T. Boothroyd.
\newblock {Spin fluctuations away from ($\ensuremath{\pi},0$) in the
  superconducting phase of molecular-intercalated FeSe}.
\newblock {\em Phys. Rev. B}, 87:220508, Jun 2013.

\bibitem{Taylor2012PRB}
A.~E. Taylor, R.~A. Ewings, T.~G. Perring, J.~S. White, P.~Babkevich,
  A.~Krzton-Maziopa, E.~Pomjakushina, K.~Conder, and A.~T. Boothroyd.
\newblock {Spin-wave excitations and superconducting resonant mode in
  Cs${}_{x}$Fe${}_{2\ensuremath{-}y}$Se${}_{2}$}.
\newblock {\em Phys. Rev. B}, 86:094528, Sep 2012.

\bibitem{nature2008Cruz}
Clarina de~la Cruz, Q.~Huang, J.~W. Lynn, Jiying Li, W.~Ratcliff, II, J.~L.
  Zarestky, H.~A. Mook, G.~F. Chen, J.~L. Luo, N.~L. Wang, and Pengcheng Dai.
\newblock {Magnetic order close to superconductivity in the iron-based layered
  LaO$_{1-x}$F$_x$FeAs systems}.
\newblock {\em {Nature}}, {453}({7197}):{899}, {JUN 12} {2008}.

\bibitem{Huang2008PRL}
Q.~Huang, Y.~Qiu, Wei Bao, M.~A. Green, J.~W. Lynn, Y.~C. Gasparovic, T.~Wu,
  G.~Wu, and X.~H. Chen.
\newblock {Neutron-Diffraction Measurements of Magnetic Order and a Structural
  Transition in the Parent ${\mathrm{BaFe}}_{2}{\mathrm{As}}_{2}$ Compound of
  FeAs-Based High-Temperature Superconductors}.
\newblock {\em Phys. Rev. Lett.}, 101:257003, Dec 2008.

\bibitem{PRB2009Li}
Shiliang Li, Clarina de~la Cruz, Q.~Huang, G.~F. Chen, T.-L. Xia, J.~L. Luo,
  N.~L. Wang, and Pengcheng Dai.
\newblock {Structural and magnetic phase transitions in
  ${\rm{Na}}_{1\ensuremath{-}\ensuremath{\delta}}\rm{FeAs}$}.
\newblock {\em Phys. Rev. B}, 80:020504, Jul 2009.

\bibitem{Hu2012PRB}
Jiangping Hu, Bao Xu, Wuming Liu, Ning-Ning Hao, and Yupeng Wang.
\newblock {Unified minimum effective model of magnetic properties of iron-based
  superconductors}.
\newblock {\em Phys. Rev. B}, 85:144403, Apr 2012.

\bibitem{Glasbrenner2015NatPhy}
J.~K. Glasbrenner, I.~I. Mazin, Harald~O. Jeschke, P.~J. Hirschfeld, R.~M.
  Fernandes, and Roser Valent{\'i}.
\newblock {Effect of magnetic frustration on nematicity and superconductivity
  in iron chalcogenides}.
\newblock {\em {Nature Physics}}, 11(11):953, Nov 2015.

\bibitem{Mizusaki2006PRB}
Takahiro Mizusaki and Masatoshi Imada.
\newblock {Gapless quantum spin liquid, stripe, and antiferromagnetic phases in
  frustrated Hubbard models in two dimensions}.
\newblock {\em Phys. Rev. B}, 74:014421, Jul 2006.

\bibitem{Imada1998RMP}
Masatoshi Imada, Atsushi Fujimori, and Yoshinori Tokura.
\newblock {Metal-insulator transitions}.
\newblock {\em Rev. Mod. Phys.}, 70:1039, Oct 1998.

\bibitem{Kyung2006PRL}
B.~Kyung and A.-M.~S. Tremblay.
\newblock {Mott Transition, Antiferromagnetism and $d$-Wave Superconductivity
  in Two-Dimensional Organic Conductors}.
\newblock {\em Phys. Rev. Lett.}, 97:046402, Jul 2006.

\bibitem{Nevidomskyy2008PRB}
Andriy~H. Nevidomskyy, Christian Scheiber, David S\'en\'echal, and A.-M.~S.
  Tremblay.
\newblock {Magnetism and $d$-wave superconductivity on the half-filled square
  lattice with frustration}.
\newblock {\em Phys. Rev. B}, 77:064427, Feb 2008.

\bibitem{Hassan2008PRB}
S.~R. Hassan, B.~Davoudi, B.~Kyung, and A.-M.~S. Tremblay.
\newblock {Conditions for magnetically induced singlet $d$-wave
  superconductivity on the square lattice}.
\newblock {\em Phys. Rev. B}, 77:094501, Mar 2008.

\bibitem{Misumi2016JPSJ}
Kazuma Misumi, Tatsuya Kaneko, and Yukinori Ohta.
\newblock {Phase Diagram of the Frustrated Square-Lattice Hubbard Model:
  Variational Cluster Approach}.
\newblock {\em J. Phys. Soc. Jpn.}, 85(6):064711, 2016.

\bibitem{Zhou2017RMP}
Yi~Zhou, Kazushi Kanoda, and Tai-Kai Ng.
\newblock {Quantum spin liquid states}.
\newblock {\em Rev. Mod. Phys.}, 89:025003, Apr 2017.

\bibitem{Misumi2017PRB}
Kazuma Misumi, Tatsuya Kaneko, and Yukinori Ohta.
\newblock {Mott transition and magnetism of the triangular-lattice Hubbard
  model with next-nearest-neighbor hopping}.
\newblock {\em Phys. Rev. B}, 95:075124, Feb 2017.

\bibitem{Tocchio2008PRB}
Luca~F. Tocchio, Federico Becca, Alberto Parola, and Sandro Sorella.
\newblock {Role of backflow correlations for the nonmagnetic phase of the
  $t{--}{t}^{\prime}$ Hubbard model}.
\newblock {\em Phys. Rev. B}, 78:041101, Jul 2008.

\bibitem{Yoshikawa2009PRB}
Toshihiko Yoshikawa and Masao Ogata.
\newblock {Role of frustration and dimensionality in the Hubbard model on the
  stacked square lattice: Variational cluster approach}.
\newblock {\em Phys. Rev. B}, 79:144429, Apr 2009.

\bibitem{Yamada2013PRB}
A.~Yamada, K.~Seki, R.~Eder, and Y.~Ohta.
\newblock {Magnetic properties and Mott transition in the square-lattice
  Hubbard model with frustration}.
\newblock {\em Phys. Rev. B}, 88:075114, Aug 2013.

\bibitem{Yamaki2013PRB}
S.~Yamaki, K.~Seki, and Y.~Ohta.
\newblock {Ground-state phase diagram of the asymmetric Hubbard model with
  geometrical frustration}.
\newblock {\em Phys. Rev. B}, 87:125112, Mar 2013.

\bibitem{Mazurenko2017Nature}
Anton Mazurenko, Christie~S. Chiu, Geoffrey Ji, Maxwell~F. Parsons, M{\'a}rton
  Kan{\'a}sz-Nagy, Richard Schmidt, Fabian Grusdt, Eugene Demler, Daniel Greif,
  and Markus Greiner.
\newblock {A cold-atom Fermi${-}$Hubbard antiferromagnet}.
\newblock {\em Nature}, 545(7655):462, May 2017.

\bibitem{Drewes2017PRL}
J.~H. Drewes, L.~A. Miller, E.~Cocchi, C.~F. Chan, N.~Wurz, M.~Gall, D.~Pertot,
  F.~Brennecke, and M.~K\"ohl.
\newblock {Antiferromagnetic Correlations in Two-Dimensional Fermionic
  Mott-Insulating and Metallic Phases}.
\newblock {\em Phys. Rev. Lett.}, 118:170401, Apr 2017.

\bibitem{Zhang2012PRB}
Yu-Zhong Zhang, Hunpyo Lee, Hai-Qing Lin, Chang-Qin Wu, Harald~O. Jeschke, and
  Roser Valent\'{\i}.
\newblock {General mechanism for orbital selective phase transitions}.
\newblock {\em Phys. Rev. B}, 85:035123, Jan 2012.

\bibitem{Yu2010PRB}
Zeng-Qiang Yu and Lan Yin.
\newblock {Collinear antiferromagnetic state in a two-dimensional Hubbard model
  at half filling}.
\newblock {\em Phys. Rev. B}, 81:195122, May 2010.

\bibitem{JPSJ.65.2559}
Hisashi Kondo and T\^{o}ru Moriya.
\newblock {On the Metal-Insulator Transition in a Two-Dimensional Hubbard
  Model}.
\newblock {\em Journal of the Physical Society of Japan}, 65(8):2559--2563,
  1996, and references thereafter.

\bibitem{Hofstetter}
W.~Hofstetter and D.~Vollhardt.
\newblock {Frustration of antiferromagnetism in the t-t'-Hubbard model at weak
  coupling}.
\newblock {\em Annalen der Physik}, 7(1):48--55, Jan 1998, and references
  thereafter.

\bibitem{Li2012JMMM}
Tie-Jun Li, Ya-Min Quan, Da-Yong Liu, and Liang-Jian Zou.
\newblock {Magnetic phase diagram of an extended Hubbard model at half filling:
  Possible application for strongly correlated iron pnictides}.
\newblock {\em {J. Magn. Magn. Mater.}}, {324}({6}):{1046}, {MAR} {2012}.

\bibitem{Qazilbash2009NatPhy}
M.~M. Qazilbash, J.~J. Hamlin, R.~E. Baumbach, Lijun Zhang, D.~J. Singh, M.~B.
  Maple, and D.~N. Basov.
\newblock {Electronic correlations in the iron pnictides}.
\newblock {\em {Natue Physics}}, {5}({9}):{647}, {SEP} {2009}.

\bibitem{Yin2011NatMat}
Z.~P. Yin, K.~Haule, and G.~Kotliar.
\newblock {Kinetic frustration and the nature of the magnetic and paramagnetic
  states in iron pnictides and iron chalcogenides}.
\newblock {\em {Nature Materials}}, {10}({12}):{932}, {DEC} {2011}.

\bibitem{JPSJ2010Miyake}
Takashi Miyake, Kazuma Nakamura, Ryotaro Arita, and Masatoshi Imada.
\newblock {Comparison of Ab initio Low-Energy Models for LaFePO, LaFeAsO,
  BaFe$_2$As$_2$, LiFeAs, FeSe, and FeTe: Electron Correlation and Covalency}.
\newblock {\em J. Phys. Soc. Jpn.}, 79(4):044705, 2010.

\bibitem{Lee2010PRB}
Hunpyo Lee, Yu-Zhong Zhang, Harald~O. Jeschke, and Roser Valent\'{\i}.
\newblock {Possible origin of the reduced ordered magnetic moment in iron
  pnictides: A dynamical mean-field theory study}.
\newblock {\em Phys. Rev. B}, 81:220506, Jun 2010.

\bibitem{Werner2016}
Philipp Werner, Shintaro Hoshino, and Hiroshi Shinaoka.
\newblock {Spin-freezing perspective on cuprates}.
\newblock {\em Phys. Rev. B}, 94:245134, Dec 2016.

\end{thebibliography}

\end{document}